\documentclass[12pt,preprint]{aastex}
\usepackage{amsmath}

\shorttitle{McNeil Nebula Outburst}
\shortauthors{McGehee et al.}

\begin{document}


\title{The V1647 Ori (IRAS 05436-0007) Protostar And Its Environment}


\author{Peregrine M. McGehee\altaffilmark{\ref{LANL},\ref{NMSU}},
J. Allyn Smith\altaffilmark{\ref{LANL2},\ref{Wyo}},
Arne A. Henden\altaffilmark{\ref{UNSO}},
Michael W. Richmond\altaffilmark{\ref{RIT}},
Gillian R. Knapp\altaffilmark{\ref{Princeton}},
Douglas P. Finkbeiner\altaffilmark{\ref{Princeton}},
\v{Z}eljko Ivezi\'{c}\altaffilmark{\ref{Princeton}},
J. Brinkmann\altaffilmark{\ref{APO}}
}
\email{peregrin@apo.nmsu.edu}

\altaffiltext{1}{Los Alamos National Laboratory, LANSCE-8,
MS H820, Los Alamos, NM 87545 
\label{LANL}}

\altaffiltext{2}{Department of Astronomy, New Mexico State University,
MSC 4500, Box 30001, Las Cruces, NM 88003 
\label{NMSU}}

\altaffiltext{3}{Los Alamos National Laboratory, ISR-4,
MS D448, Los Alamos, NM 87545 
\label{LANL2}}

\altaffiltext{4}{Department of Physics \& Astronomy, 
University of Wyoming, 1000 E. University Blvd.,  Laramie, WY 82071 
\label{Wyo}}

\altaffiltext{5}{Universities Space Research Association/US Naval Observatory, 
Flagstaff Station,
P.O. Box 1149, Flagstaff, AZ 86002 
\label{UNSO}}

\altaffiltext{6}{Department of Physics, 
Rochester Institute of Technology,
85 Lomb Memorial Drive, Rochester, NY 14623
\label{RIT}}

\altaffiltext{7}{Princeton University Observatory, Princeton, NJ 08544
\label{Princeton}}

\altaffiltext{8}{Apache Point Observatory, 2001 Apache Point Road, Sunspot, NM 88349
\label{APO}}


\begin{abstract}
We present Sloan Digital Sky Survey and United States Naval Observatory
observations of the V1647 Ori protostar and surrounding field near 
NGC 2068. V1647 Ori, the likely driving source for HH 23, brightened 
significantly in November 2003.
Analysis of SDSS imaging acquired in November 1998 and February 
2002 during the quiescent state, recent USNO photometry, and published 
2MASS and Gemini data shows that the color changes associated with 
brightening suggest an EXor outburst rather than a simple dust 
clearing event.
\end{abstract}


\keywords{stars: formation - stars: pre-main-sequence -
stars: circumstellar matter - stars: individual (V1647 Ori, IRAS 05436-0007)}

\section{Introduction}

In January 2004, J.W. McNeil discovered a new reflection nebula in the dark 
cloud Lynds 1630 near M 78 \citep{mcn04} This object,
now known as McNeil's Nebula, is apparently associated with an EXor-type
eruption \citep{rei04} of the embedded protostar V1647 Ori.

EXors belong to the class of pre-main-sequence optical variables \citep{her77}.
They are classical T Tau stars which undergo irregular outbursts
in the optical/UV of several magnitudes, named for the prototype EX Lup
\citep{her01}.  These outbursts are interpreted as episodes of
substantial mass transfer resulting from instabilities in the accretion
disk; they are present very early in the evolution of a protostar, as 
shown by the detection of EXor outbursts from deeply embedded Class 1
protostars in the Serpens star formation region \citep{hod96}.

\citet{cla91} first identified V1647 Ori as the young stellar object
IRAS 05436-0007 on the 
basis of its IRAS colors. I-band and [SII] narrow-band imaging of the
region by \citet{eis97} revealed a faint I band source
at the position of the IRAS object and reflection nebulosity extending to
the north, identifying V1647 Ori as the likely driver for HH 23, located
170 arcsec north of the star. 
The bolometric flux of the source derived from IRAS and sub-millimeter
photometry by \citet{lis99} yields a luminosity of
2.7 $\rm L_{\odot}$ and an inferred molecular gas mass of 0.4 $\rm
M_{\odot}$ assuming a distance of 400 pc to the Orion star-forming complex
\citep{ant82}.

Estimates of the extinction towards V1647 Ori,
$\rm A_V$ = $11^m$ - $15^m$, are found to be similar 
from photometry taken during the
quiescent phase (Abraham et al. 2004)
and during the eruptive phase (Reipurth \& Aspin 2004; Vacca et al. 2004;
Brice\~no et al. 2004; Andrews, Rothberg \& Simon 2004).  Thus, 
it is not clear whether the appearance
of McNeil's Nebula is due only to the eruption of V1647 Ori or to the 
eruption plus
additional clearing of obscuring circumstellar dust.
In this paper, we examine this question
using pre-eruption multiband optical and near infrared data from the Sloan 
Digital Sky Survey (SDSS) and the Two Micron All Sky Survey (2MASS)
compared with post-eruption data in the SDSS and 2MASS bands observed at
the United States Naval Observatory.
  
\section{Observations}

We detect the protostar at four epochs of Sloan Digital Sky Survey (SDSS)
imaging as a point source (SDSS J054613.14-000604.1) coincident with 
the 2MASS $K$-band 
position ($\alpha_{2000}$=05$^h$46$^m$13.1$^s$, 
$\delta_{2000}$=-00$^o$06$^{'}$05$^{''}$). 
The SDSS observations consist of two pairs of scans acquired in November 
1998 and February 2002.
Figure \ref{fig-detail} shows an SDSS composite image made with the $riz$ 
filters in which the protostar and the faint nebulosity to the north can
be seen.

A technical summary of the SDSS is given by \citet{yor00}.
The SDSS imaging camera is described by \citet{gun98}.
The Early Data Release and the Data Release One are described by
\citet{sto02} and \citet{aba03}. The former includes an extensive 
discussion of the data outputs and software.  \citet{pie03} describe the 
astrometric calibration of the survey and the 
network of primary photometric standard stars is described by 
\citet{smi02}. The photometric system itself is defined by \citet{fuk96},
and the system which monitors the site photometricity by
\citet{hog01}.
\citet{aba03} discuss the differences between the native SDSS 2.5m
$ugriz$ system and the $u'g'r'i'z'$ standard star system defined
on the USNO 1.0 m \citep{smi02}.

The SDSS low Galactic latitude data which includes the Orion equatorial 
imaging used in this work are described by \citet{fin04}.

  The U.S. Naval Observatory Flagstaff Station 1.0 m and 1.55 m telescopes
were used to obtain eruptive phase $BVRIz'$ and $JHK$ photometry of 
V1647 Ori.
For $BVRIz'$, frames were taken and flatfielded using twilight sky flats.
DAOPHOT PSF fitting as implemented in IRAF was used to obtain
photometric measures of the target since there is a bright part
of the nebula only a few arcsec distant.  For each dataset,
ensemble differential photometry was performed using a set of
secondary standard stars calibrated on two photometric nights
with the USNO 1.0 m telescope.  The $z'$ measures were relative to
$z'$ secondary standard stars calibrated by the SDSS PT telescope
\citep{hog01}.
For $JHK$, we used 2MASS stars in the field, eliminating obvious
variables, to calibrate the data.  
We used a standard $K$ filter for 2004 February 11, but used a
$K'$ filter for 2004 April 12.

\section{Results}

\begin{figure}
\plotone{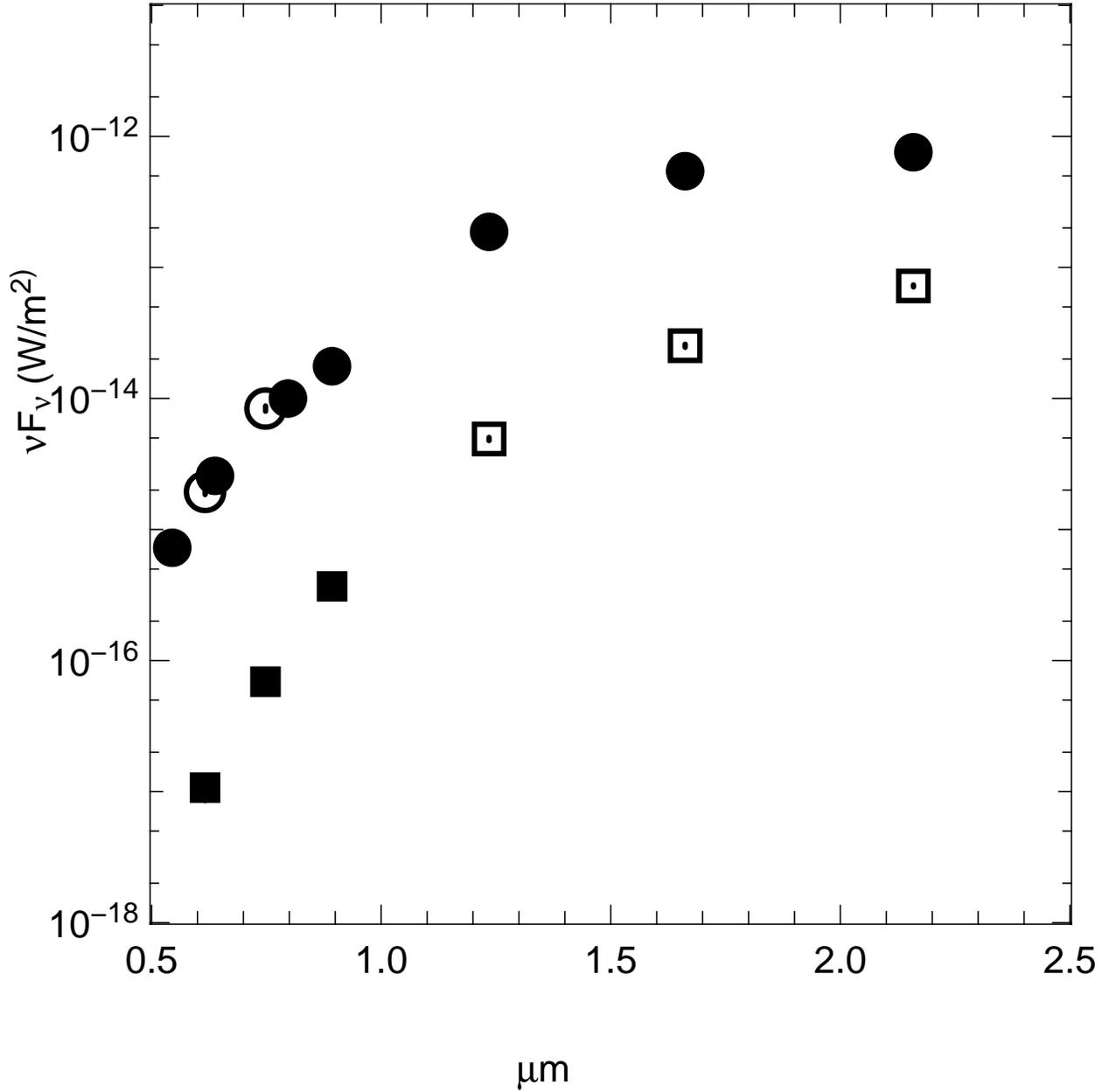}
\caption{{\bf Quiescent and Eruptive Spectral Energy Distributions.}
The optical and near-IR SEDs are shown for the quiescent state ({\it squares})
and during eruption ({\it circles}). Data acquired by the SDSS and at the
USNO are indicated by filled symbols. The open squares are the 2MASS $JHK$
observations and the open circles are the $r$ and $i$ band eruptive 
phase measurements
of \citet{rei04}.
\label{fig-sed}}
\end{figure}

The pre-eruptive and recent optical and near-infrared photometry of
V1647 Ori are summarized in Tables \ref{tbl-photo_opt},
\ref{tbl-photo_erupt}, and \ref{tbl-photo_nir}. The quiescent and
eruptive phase spectral energy distributions are shown as
$\nu F_\nu$ in Figure \ref{fig-sed}. 

We examine the evolution of reddening invariant colors from the
quiescent to the eruptive states to constrain the underlying
physical processes. These colors take the generic form
$\rm C_{xyz} = (x-y) - (y-z) * E(x-y)/E(y-z)$ where $x, y, $ and $z$ are
the observed magnitudes in each passband. 
A color change
is computed as $\Delta \rm C_{xyz}$,
where color changes having $\Delta \rm C_{xyz}$ statistically distinct 
from 0 indicate changes in the spectral energy distribution (SED)
not consistent with pure dust clearing.
The conversion from $\rm E(B-V)$ to extinction in each band follows
\citet{sch98} and Finkbeiner et al.(2004), {\it in preparation}

The reddening invariant
colors measured before and during the eruption are listed in Table
\ref{tbl-colors} for both $\rm R_V = 3.1$ and
$\rm R_V = 5.5$, the latter appropriate for the larger dust grains found
in star formation regions. $\rm R_V$, defined as $\rm A_V/E(B-V)$, is the 
ratio of the general to selective extinction.

Characteristic grain sizes are inferred
to increase from 0.17 $\mu$m to 0.21 $\mu$m as $\rm R_V$ ranges from 3.1 to
5.5 assuming a mix of silicate and carbonaceous populations \citep{wei01}. 
In the coldest
portions of the molecular clouds additional grain species such
as refractory and volatile organics, olivine, water ice, orthopyroxene, 
trolite, and metallic iron are expected to contribute to the opacity
\citep{pol94}. \citet{vac04} detect absorption due to water ice at
3.1 $\mu$m in the near-IR outburst spectrum.

Study of the selective extinction in background stars has revealed that
$\rm R_V$ is not constant within the interiors of dark molecular clouds. 
\citet{str01} find that for the Bok globule CB 107
$\rm R_V$ increases inwards reaching a value of 6.5 at the core. Given
the complex nature of the V1647 Ori environment including
outburst cavities, accretion disks, and remnant envelopes it is
likely that $\rm R_V$ can vary on small spatial scales. 

\subsection{Determination of Extinction}

In the foregoing analysis we choose to characterize the intervening
dust in terms of the SDSS $z$ band extinction and $\rm R_V$. Since the shape of
the extinction curve is dependent upon the grain size distribution
for $\lambda < 0.9 \mu$m \citep{car89} the traditional
$\rm A_V$ and $\rm E(B-V)$ quantities are sensitive to both the dust 
column density and $\rm R_V$. By adopting reference passbands longward 
of 0.9 $\mu$m we remove the dependency on the form of the dust law 
although $\rm R_V$ must still be considered in the optical and UV.
Alternate passbands that are free of $\rm R_V$ effects include the near-IR
$J$ and Johnson-Cousins $I$, the latter used by \citet{wei01} who
adopted $\rm A_I/N(H I) = 2.5 \times 10^{-22}$ cm$^2$ for a standard
gas-to-dust ratio.

Following \citet{abr04} and \citet{rei04} we assume that the
quiescent phase near-IR colors are that of an embedded low-mass 
Classical T Tauri and obtain $\rm E(J-H)$ = 1.43$^m$ by dereddening
onto the Classical T Tauri locus of \citet{mey97}.
This value of $\rm E(J-H)$ corresponds to $\rm A_J$ = 3.39$^m$ 
($\rm A_z$ = 6.40$^m$)
given $\lambda_{\rm eff}(J) = 1.25 \mu$m and 
$\lambda_{\rm eff}(H) = 1.65 \mu$m
and application of the \citet{car89} methodology. 
 
\subsection{Changes in the Spectral Energy Distribution}

In order to analyze the general differences between the quiescent
and eruptive appearance we combine sets of observations to
form representative values. The quiescent state is taken as
the October 7 (2MASS $J,H,K$) and the 1998 November 17
(SDSS $r,i,z$) observations. The optical photometry for the eruptive
state is obtained from the 2004 February 14
Gemini $r$ and $i$ data \citep{rei04}, and the 2004 February 26 USNO $z$ band
observations. The $J, H,$ and $K$ data for the eruptive state 
are from the 2004 February 11 USNO
observations.

Table \ref{tbl-deltas} presents the reddening invariant colors
in the quiescent and eruptive states for $\rm R_V$ = 3.1
and $\rm R_V$ = 5.5. As evidenced by the greater than 5$\sigma$ change
in all colors except for $\rm C_{riz}$, which has a marginal $r$ band 
detection in the quiescent state,
it is clear that the increase in emission is not
consistent with a
dust clearing event and thus we conclude that an intrinsic
change occurred to the SED of V1647 Ori.

In Figure \ref{fig-delm} we compare the observed magnitude 
variations
against $\Delta \rm A_z$ = $-2^m$ and $-4^m$ dust clearing events and, as 
concluded above based on reddening invariant colors, find they are not 
explainable in terms of a simple diminishing of the line of sight 
extinction. 

\subsection{The Quiescent Phase}

From observations of Class II protostars (Classical T Tauris)
we would expect that the veiling continuum due to magnetospheric accretion
will be present in the SDSS $u$ and $g$ bands but weakening into the redder
$r$, $i$ and $z$ bands \citep{cal98}. In a similar fashion,
the thermal emission
from the inner portion of the circumstellar disk should be
bright in the $H$ and $K$ bands but diminishing into the bluer $J$ band
 \citep{mey97}. For all but the mostly heavily veiled stars the
$i$ and $z$ bands will only be affected by extinction. 
None of the reddening invariant colors can be used to give the
intrinsic spectral type.

Young (1 Myr) T Tauri stars exhibit spectral types between 
M0 and M4 for masses 
of 0.1 to 1.0 M$_\odot$ \citep{bar98}. The use of the \citet{mey97} near-IR 
Classical T Tauri
locus by \citet{abr04} and \citet{rei04} implicitly assumes a similar spectral
type as this locus is based on
observations of
K7/M0 stars in the Taurus star formation region.
The $i-z$ colors for these early to mid M spectral
types range from 0.38 to 0.80 with considerable scatter on the order
of several tenths of magnitudes \citep{wes04}. Given the observed $i-z$
(2.01$\pm$0.06), dereddening to these intrinsic colors requires $\rm A_z$ =
3.1$^m$ to 4.2$^m$ for $\rm R_V=3.1$ and 4.3$^m$ to 5.7$^m$ for 
$\rm R_V = 5.5$. 

Dereddening using the estimated $\rm A_z \sim 6.4$ and $\rm R_V$ = 3.1 
results in an intrinsic $i-z$ of -0.47, which is too blue for the stellar 
locus. Assumption of $\rm R_V$ = 5.5 yields an $i-z$ of 0.20, corresponding
to a late K spectral type \citep{fin00}. 
If V1647 Ori is indeed a low mass protostar with a late K or M spectral type 
than either $R_V$ is high or significant $i$ band veiling is present 
in the quiescent state. 

\begin{figure}
\plotone{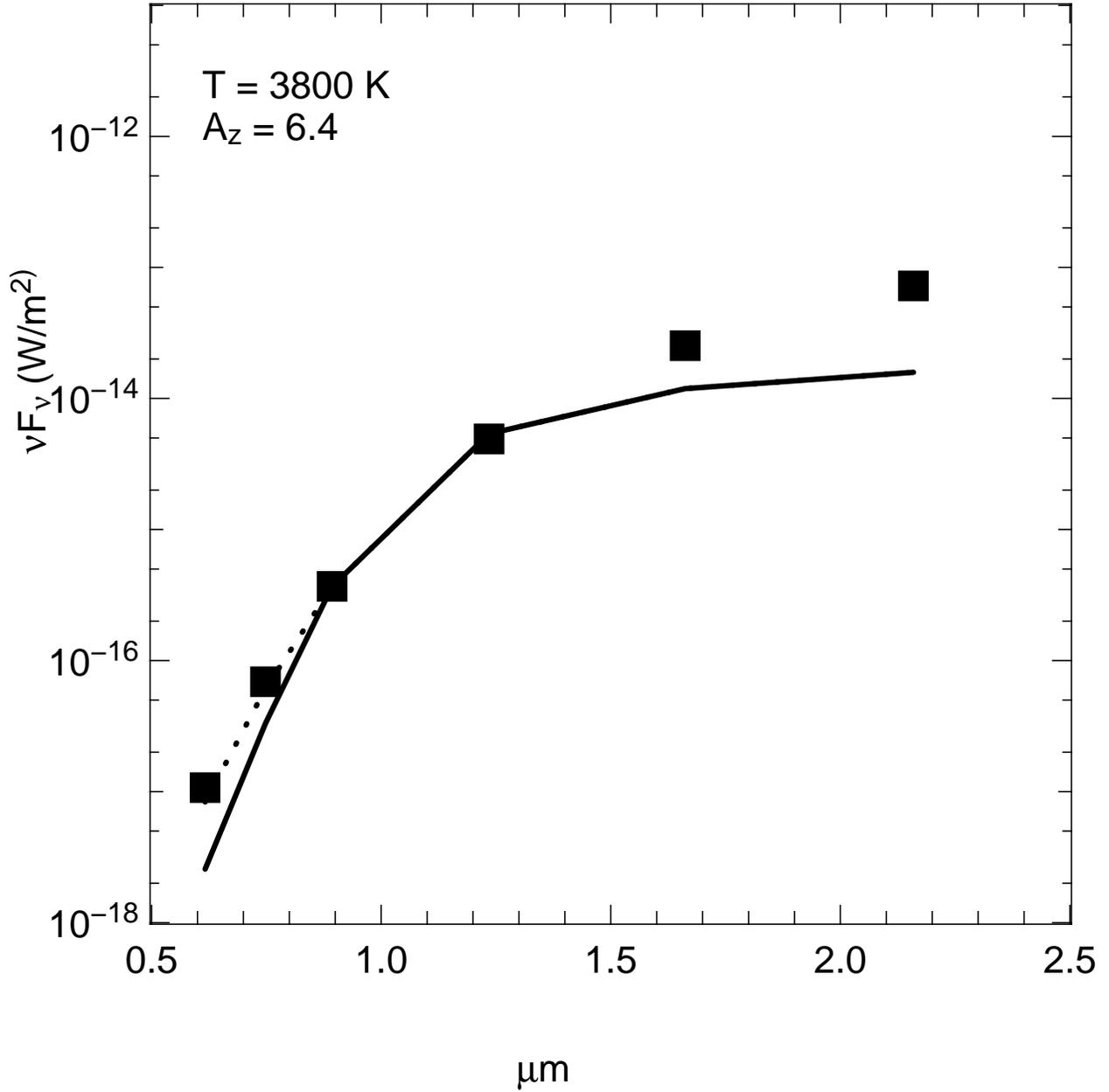}
\caption{{\bf Modeling the Quiescent Spectral Energy Distribution.}
Observed pre-eruption SED compared with an
M0 photosphere (3800 K) seen under a line of sight
reddening of $\rm A_z$ = 6.4$^m$ for $\rm R_V$ = 3.1 ({\it solid line}) 
and $\rm R_V$ = 5.5 ({\it dotted
line}). Excess emission is evident longward of 1.2 $\mu$m ($J$).
\label{fig-qsed}}
\end{figure}

In Figure \ref{fig-qsed} we compare the observed quiescent phase
SED against an M0 photosphere seen under an extinction of $\rm A_z$=6.4
magnitudes. Excess emission is seen in the $J, H,$ and $K$ bands
that could indicate the presence of a circumstellar disk.

\subsection{The Eruptive Phase}

\begin{figure}
\plotone{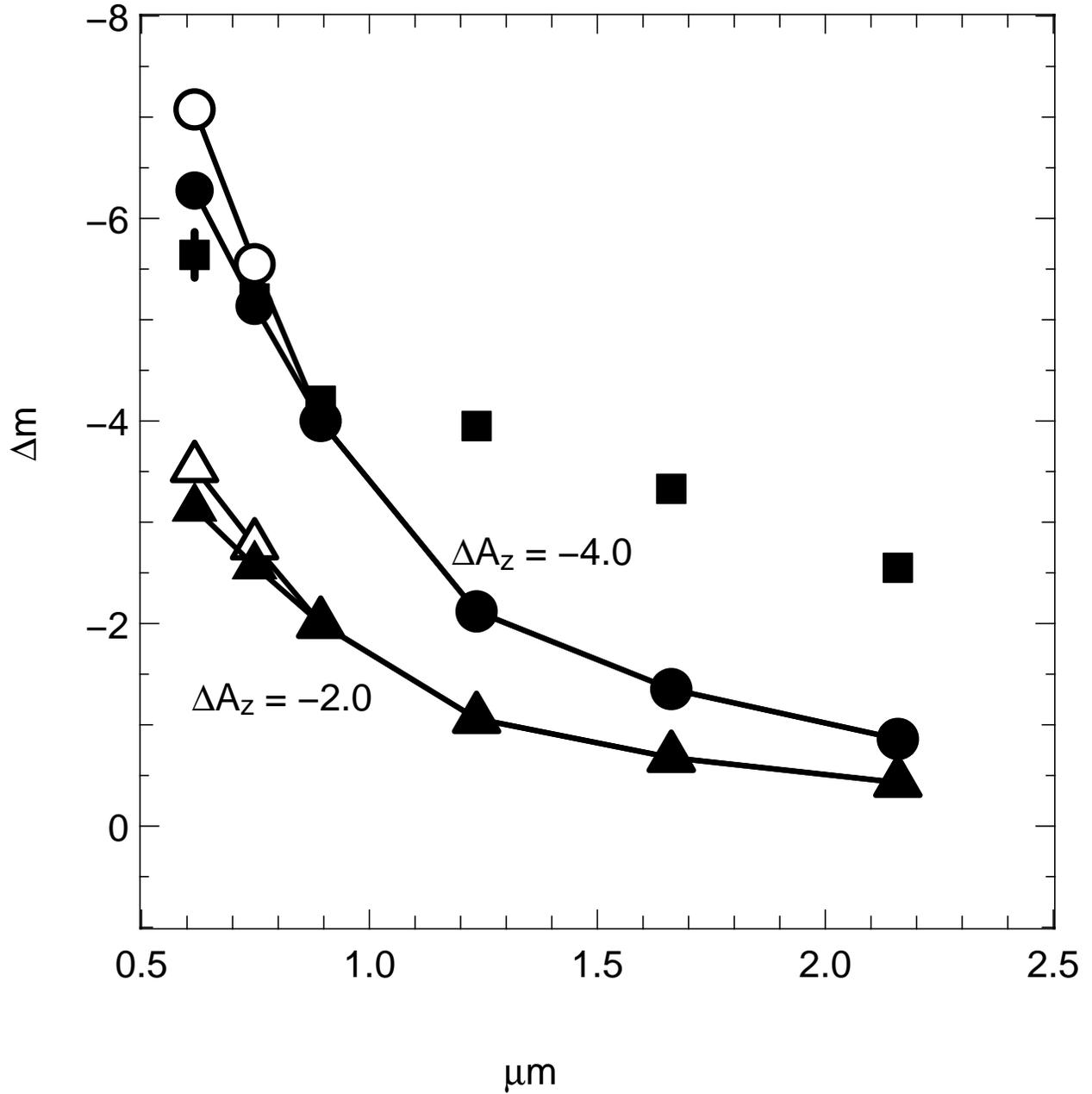}
\caption{{\bf Changes in Spectral Energy Distributions.}
The observed magnitude differences ({\it squares}) are compared 
against pure dust clearing events for $\Delta \rm A_z$ = -2$^m$ ({\it circles})
and $\Delta \rm A_z$ = -4$^m$ ({\it triangles}) and 
for $\rm R_V$ = 3.1 ({\it solid}) or $\rm R_V$ = 5.5 ({\it open}) dust laws.
\label{fig-delm}}
\end{figure}

The optical spectra acquired during the outburst by \citet{wal04}
lack the TiO molecular absorption bands characteristic
of late K and M stars. \citet{wal04} attribute this to either
an early photospheric spectral type or overwhelming veiling. We adopt
the latter interpretation given the apparent change to the intrinsic
SED and the large brightness increase in the optical (Figure \ref{fig-delm}).

If the new component to the SED is due to an EXor outburst then we 
expect to see
emission from the high temperature (6000 - 8000 K) inner disk
\citep{bel95}. In  Figure \ref{fig-delf} we show the observed flux
($\nu F_\nu$) increase in comparison with a 7000 K blackbody 
reddened by $\rm A_z$ = 3.2$^m$ and 6.4$^m$
for $\rm R_V$ = 3.1 and 5.5. In common with \citet{wal04} we see that a 
single EXor-like high temperature component can not reproduce the 
observed excess in both the optical and the near-IR.

For the partial dust clearing events suggested by \citet{rei04} 
and \citet{wal04} we find that due to the increase in optical depth
towards the blue the additional flux due to a new source must
fall more rapidly at the shorter wavelengths. The observed change in
$J-H$ is -0.61$^m$, which if entirely due to dust clearing requires
$\Delta \rm A_z$ = -3.2$^m$. The intrinsic flux increase required in
this case is shown in Figure \ref{fig-delf} and 
would require either a lower temperature for the new source
or a significantly higher extinction than anticipated.

Spectral indices are commonly defined by the ratio of the
flux of the feature ($F_s$) and that of the nearby pseudo-continuum
($F_c$), so the intrinsic measure of the feature strength is
$I_0 = F_s/F_c$ \citep{giz97}.
If a veiling continuum is present, defined by
$F_v = r F_c$, then the measured spectral index, $I$, is

\begin{equation}
\label{eqn-veil}
\begin{split}
I & = \frac{F_s + r F_c}{F_c + r  F_c} \\
  & = \frac{I_0}{1 + r} + \frac{r}{1 + r}. \\
\end{split}
\end{equation}

\begin{figure}
\plotone{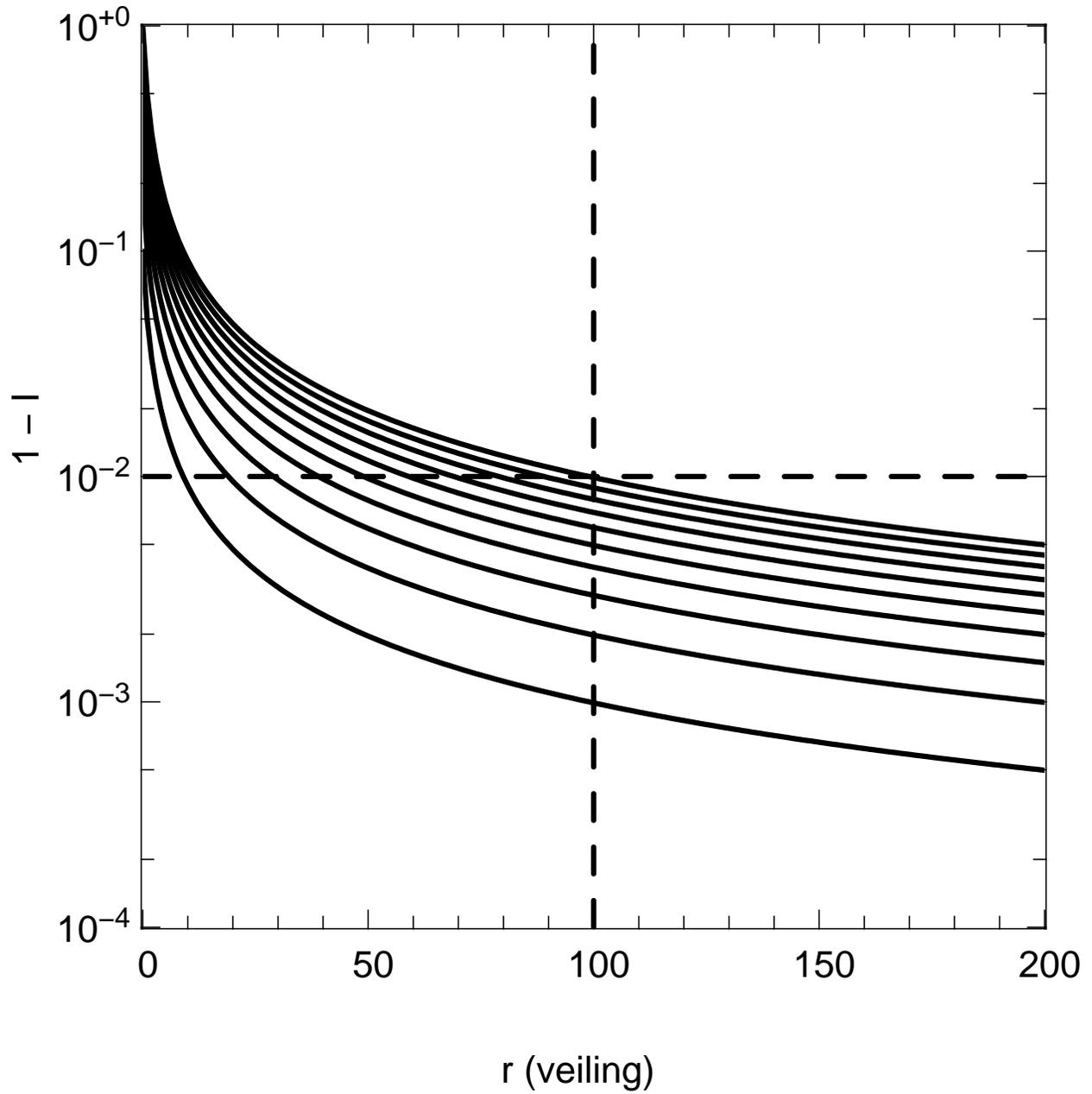}
\caption{{\bf Effect of Veiling on Spectral Features.}
This figure plots the observed depression of an absorption line 
relative to the continuum level $(1 - I)$ on a logarithmic scale as a 
function of the veiling, $r$, for unveiled indices of $I_0$ = 0.0, 0.1, to 0.9.
The dashed lines indicate the eruptive phase $i$ band veiling of 100 and
a relative depression level of 1\%.
\label{fig-veil}}
\end{figure}

Figure \ref{fig-veil} shows the
observed depression of the line relative to the continuum level
$(1 - I)$ on a logarithmic scale as a function of the veiling, $r$,
for unveiled indices of $I_0$ = 0.0, 0.1, to 0.9.
From Figure \ref{fig-delm} we see that the flux in the $i$ band has 
increased by $\Delta$m = -5 or a factor of $\sim$100 which means 
the TiO features will be visible as less than a 1\% fluctuation
in the observed continuum.
  
\section{Conclusions}

The observed photometric variations are not consistent with a simple
dust clearing event as evidenced by changes in reddening invariant indices.
We infer that the process
of eruption involves changes to both the optical and near-IR SED.

Application of an inferred pre-outburst extinction of $\rm A_z$ = 6.4$^m$
suggests a late K spectral type 
if the $r-i$ and $i-z$ colors
are minimally affected by the veiling continuum arising from magnetospheric 
accretion shocks and that $R_V$ has a value expected for a star formation
region. An early to mid M spectral type is possible if the $i$ band 
includes an excess emission of 0.2$^m$ to 0.6$^m$, corresponding to
a veiling between 0.2 and 0.7.

We interpret the quiescent phase of V1647 Ori as an
embedded low mass Classical T Tauri.
Comparison with a reddened M0 photosphere 
shows a near-IR excess that is presumably due to thermal emission
from a normal circumstellar disk. 

\begin{figure}
\plotone{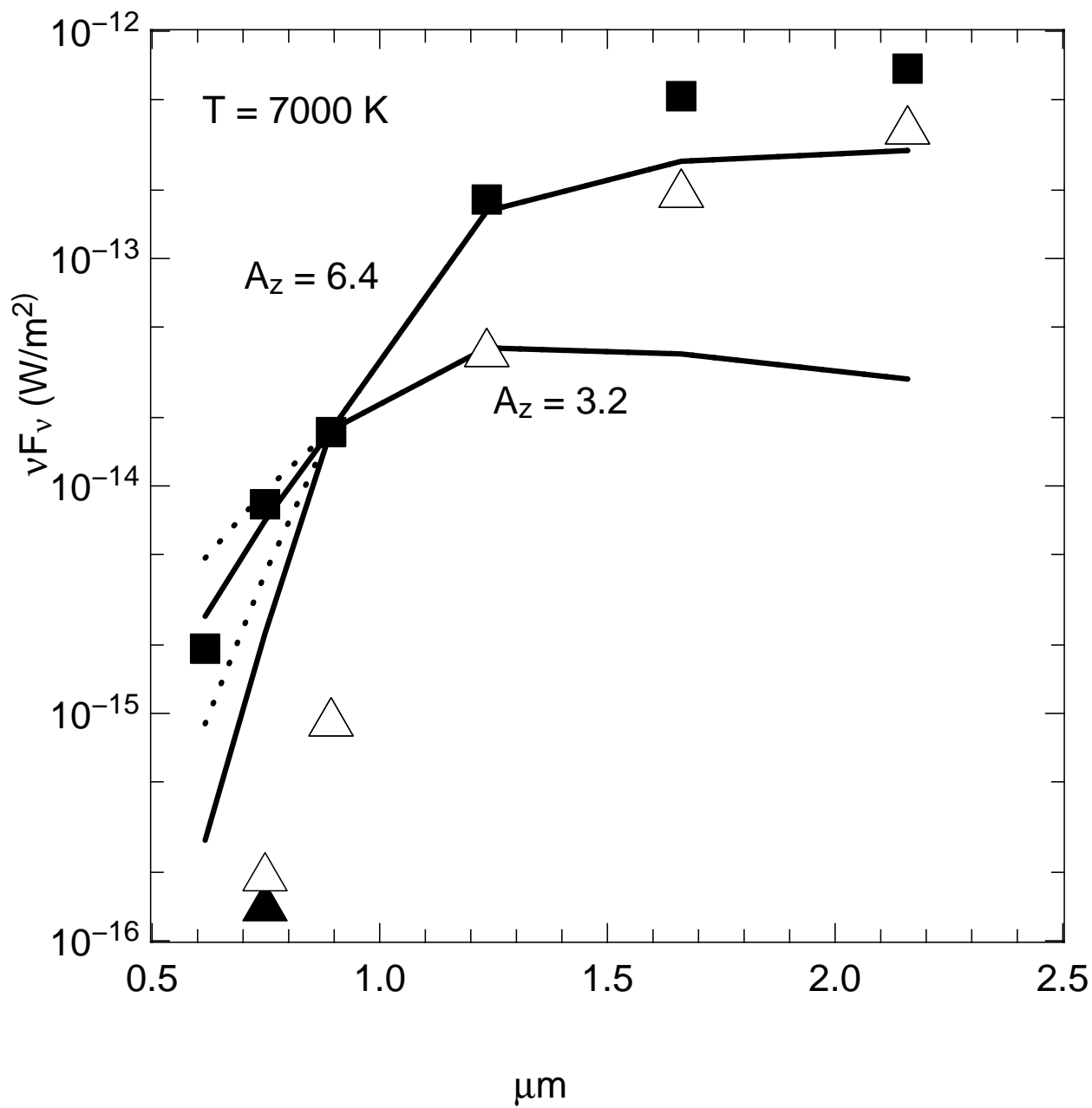}
\caption{{\bf Modeling the Eruptive Component.}
This figure compares the observed flux increase 
{(\it filled squares}) against 
a 7000 K blackbody
viewed under ($\rm R_V$=3.1 ({\it solid}), 5.5 ({\it dotted})) 
extinctions of $\rm A_z$ = 3.2$^m$ and
6.4$^m$,
scaled to match the observed change in the $z$ band flux. 
The triangles show the required intrinsic flux increase in the 
case of a $\Delta \rm A_z = -3.2^m$ partial dust clearing for
$\rm R_V$ = 3.1 ({\it filled}) and 5.5 ({\it open}).
\label{fig-delf}}
\end{figure}

The outburst SED is dominated by the new component. We see that
a single blackbody having the $\sim$7000 K temperature expected for an 
EXor inner disk can not simultaneously reproduce both the optical
and near-IR portions of the SED.
It is possible, as suggested by \citet{rei04}, that a partial
dust clearing event occurred in combination with an intrinsic brightening.
We note that a reduction of the line of sight extinction would
preferentially lessen the required flux increase at shorter wavelengths.
This steepening of the curve shown in Figure \ref{fig-delf} would increase the
difficulty of fitting the high temperature (B spectral type)
component observed by \citet{bri04}.

Further study of V1647 Ori planned using the facilities at USNO, Apache 
Point Observatory, and the Spitzer Space Telescope may clarify the
nature and evolutionary state of this young star.

\acknowledgments

Funding for the creation and distribution of the SDSS Archive has been 
provided by the Alfred P. Sloan Foundation, the Participating 
Institutions, the National Aeronautics and Space Administration, the 
National Science Foundation, the U.S. Department of Energy, the Japanese 
Monbukagakusho, and the Max Planck Society. The SDSS Web site is 
http://www.sdss.org/.

The SDSS is managed by the Astrophysical Research Consortium (ARC) for 
the Participating Institutions. The Participating Institutions are The 
University of Chicago, Fermilab, the Institute for Advanced Study, the 
Japan Participation Group, The Johns Hopkins University, Los Alamos 
National Laboratory, the Max-Planck-Institute for Astronomy (MPIA), the 
Max-Planck-Institute for Astrophysics (MPA), New Mexico State University, 
University of Pittsburgh, Princeton University, the United States Naval 
Observatory, and the University of Washington.

This publication makes use of data products from the Two Micron
All Sky Survey, which is a joint project of the University of
Massachusetts and the Infrared Processing and Analysis 
Center/California Institute of Technology, funded by the 
National Aeronautics and Space Administration and the National
Science Foundation.

Finally we thank the referee, Colin Aspin, for his helpful comments.


%% copy editing.  Individual instruments can be provided in parentheses,
%% after the keyword, but they will not be verified.

Facilities: \facility{SDSS}, \facility{USNO}, \facility{Gemini-N}, \facility{2MASS}.


\clearpage
\begin{deluxetable}{lrrrl}
\tablecaption{$riz$ Photometry of V1647 Ori
\label{tbl-photo_opt}}
\tablehead{
\colhead{Date} &
\colhead{$r$} &
\colhead{$i$} &
\colhead{$z$} &
\colhead{Telescope}
}
\startdata
1998 Nov 17 &
$23.04\pm0.22$ &
$20.81\pm0.05$ &
$18.80\pm0.04$ & 
SDSS \\
1998 Nov 28 &
$24.00\pm0.80$ &
$21.03\pm0.09$ &
$19.19\pm0.08$ &
SDSS \\
2002 Feb 07 &
$22.69\pm0.29$ &
$20.77\pm0.08$ &
$18.77\pm0.06$ & 
SDSS \\
2002 Feb 09 &
$23.88\pm0.73$ &
$21.33\pm0.11$ &
$19.33\pm0.07$ & 
SDSS \\
2004 Feb 14 &
17.4 & 
15.6 &
\ldots &
Gemini\tablenotemark{a} \\ 
2004 Feb 26 &
\ldots & 
\ldots & 
$14.60\pm0.06$ &
USNO 1.55m \\
2004 Apr 16 &
\ldots & 
\ldots &  
$14.79\pm0.06$ &
USNO 1.55m \\
2004 Apr 27 &
\ldots & 
\ldots & 
$14.72\pm0.09$&
USNO 1.55m \\
\enddata
\tablenotetext{a}{
From \citet{rei04}.}
\end{deluxetable}

\clearpage
\begin{deluxetable}{lrrrrl}
\tablecaption{Johnson-Cousins Photometry of V1647 Ori
\label{tbl-photo_erupt}}
\tablehead{
\colhead{Date} &
\colhead{$B$} &
\colhead{$V$} &
\colhead{$R_c$} &
\colhead{$I_c$} &
\colhead{Telescope}
}
\startdata
2004 Feb 11 &
\ldots &
$18.52\pm0.07$ &
$16.91\pm0.08$ &
$14.92\pm0.05$ &
USNO 1.0m \\
2004 Feb 16 &
$20.39\pm0.12$ &
$18.59\pm0.06$ &
$16.87\pm0.07$ &
$14.90\pm0.04$ &
USNO 1.0m \\
2004 Feb 25 &
\ldots &
$18.68\pm0.04$ &
\ldots & 
\ldots & 
USNO 1.55m \\
2004 Feb 26 &
$20.84\pm0.08$ &
$18.99\pm0.04$ &
\ldots &
$15.15\pm0.02$ &
USNO 1.55m \\
2004 Apr 13 &
$20.74\pm0.33$ &
$19.12\pm0.08$ &
$17.24\pm0.09$ &
$15.28\pm0.05$ &
USNO 1.0m \\
2004 Apr 16 &
$21.13\pm0.07$ &
$19.26\pm0.03$ &
\ldots &
$15.37\pm0.01$ &
USNO 1.55m \\
\enddata
\end{deluxetable}

\clearpage
\begin{deluxetable}{lrrrl}
\tablecaption{Near-IR Photometry of V1647 Ori
\label{tbl-photo_nir}}
\tablehead{
\colhead{Date} &
\colhead{$J$} &
\colhead{$H$} &
\colhead{$K$} &
\colhead{Telescope}
}
\startdata
1998 Oct 07 &
$14.74\pm0.03$ &
$12.16\pm0.03$ &
$10.27\pm0.02$ &
2MASS \\
2004 Feb 03 &
$11.1\pm0.1$ &
$9.0\pm0.1$ &
$7.4\pm0.1$ &
Gemini\tablenotemark{a} \\
2004 Feb 11 &
$10.79\pm0.01$ &
$8.83\pm0.01$ &
$7.72\pm0.01$\tablenotemark{b} &
USNO 1.55m \\
2004 Apr 12 &
$10.94\pm0.02$ &
$9.06\pm0.02$ &
$7.59\pm0.02$\tablenotemark{b}&
USNO 1.55m \\
\enddata
\tablenotetext{a}{
From \citet{rei04}.}
\tablenotetext{b}{
The USNO 1.55m observations used standard $K$ on 2004 February 11
and $K'$ on 2004 April 12, both placed
on the 2MASS zeropoint system, but probably with transformation
differences.}
\end{deluxetable}

\clearpage
\begin{deluxetable}{lll}
\tablecaption{Reddening Invariant Colors
\label{tbl-colors}}
\tablehead{
\colhead{Color} &
\colhead{$\rm R_V$ = 3.1} &
\colhead{$\rm R_V$ = 5.5}
}
\startdata
$C_{riz}$ &
$(r-i) - 0.987*(i-z)$ &
$(r-i) - 1.002*(i-z)$ \\
$C_{izJ}$ &
$(i-z) - 0.824*(z-J)$ &
$(i-z) - 0.605*(z-J)$ \\
$C_{zJH}$ &
$(z-J) - 2.443*(J-H)$ &
$(z-J) - 2.443*(J-H)$ \\
$C_{JHK}$ &
$(J-H) - 1.563*(H-K)$ &
$(J-H) - 1.563*(H-K)$
\enddata
\end{deluxetable}

\clearpage
\begin{deluxetable}{lrrrrrrr}
\tablecaption{Reddening Invariant Color Changes
\label{tbl-deltas}}
\tablehead{
\colhead{} &
\multicolumn{3}{c}{$\rm R_V = 3.1$} &
\colhead{} &
\multicolumn{3}{c}{$\rm R_V = 5.5$} \\
\cline{2-4} \cline{6-8} \\
\colhead{Color} &
\colhead{Quiescent\tablenotemark{a}} &
\colhead{Eruptive\tablenotemark{b}} &
\colhead{N$\sigma$\tablenotemark{c}} &&
\colhead{Quiescent\tablenotemark{a}} &
\colhead{Eruptive\tablenotemark{b}} &
\colhead{N$\sigma$\tablenotemark{c}}
}
\startdata
$\rm C_{riz}$ & 0.25$\pm$0.23 & 0.81$\pm$0.10 & 2.2 && 
0.22$\pm$0.23 & 0.80$\pm$0.11 & 2.3 \\
$\rm C_{izJ}$ & -1.34$\pm$0.08 & -2.14$\pm$0.09 & -6.7 &&
-0.45$\pm$0.07 & -1.31$\pm$0.09 & -7.7 \\
$\rm C_{zJH}$ & -2.24$\pm$0.12 & -0.98$\pm$0.07 & 9.4 &&
-2.24$\pm$0.12 & -0.98$\pm$0.07 & 9.4 \\
$\rm C_{JHK}$ & -0.37$\pm$0.07 & 0.23$\pm$0.03 & 8.0 &&
-0.37$\pm$0.07 & 0.23$\pm$0.03 & 8.0
\enddata
\tablenotetext{a}{The quiescent
state is defined by the 1998 October 7 2MASS and 1998 November 17
SDSS data.}
\tablenotetext{b}{The eruptive phase data are from the 
2004 February 14 Gemini $r$ and $i$, the 2004 February 26 USNO $z$, and the
2004 February 11 USNO $J, H,$ and $K$ measurements.}
\tablenotetext{c}{N$\sigma$ is the ratio of color change to the
quadrature sum of measurement errors.}
\end{deluxetable}

\clearpage
\begin{figure}
%%\plotone{f1COLOR.eps}
\caption{{\bf SDSS pre-eruption $riz$ band mosaic image of V1647 Ori.}
The location of V1647 Ori is marked on this 2x2 binned mosaic 
which maps the SDSS $r$, $i$, and $z$ bands
onto blue, green, and red. Herbig-Haro objects, such as the large HH 24 
complex ({\it bottom center}), are seen in blue
due to the H$\alpha$ emission appearing in the $r$ band. The
image is roughly 10 arcminutes on a side and is displayed using a
negated grayscale. North is up and east is to
the left. The faint emission immediately north of V1647 Ori
significantly brightens during the eruptive phase where it is 
seen as McNeil's Nebula. The diffuse $r$ band emission due to HH 23, which 
may be driven by V1647 Ori, is highlighted by the box in the upper center.
{\it A color rendition of this image is available in the electronic version
of the journal.}
\label{fig-detail}}
\end{figure}

\clearpage

\end{document}